\documentclass[aps,prc,showpacs]{revtex4}
\usepackage{graphicx}
\usepackage{latexsym}
\usepackage{bm}
\usepackage{dcolumn}
\usepackage{amsmath}

\newcommand{\bra}{\langle}
\newcommand{\ket}{\rangle}

\def\slass#1{#1 \hspace{-1.9mm} \slash}
\def\slasss#1{#1 \hspace{-2.3mm} \slash}

\begin{document}

\bibliographystyle{apsrev}

\title{Unitarity and the Bethe-Salpeter Equation}
\author{A. D. Lahiff}
\affiliation{TRIUMF, 4004 Wesbrook Mall, Vancouver, British Columbia,
Canada V6T 2A3}
\author{I. R. Afnan}
\affiliation{School of Chemistry, Physics, and Earth Sciences,\\
Flinders University, GPO Box 2100, Adelaide 5001, Australia}

\date{\today}
\begin{abstract}
We investigate the relation between different three-dimensional
reductions of the Bethe-Salpeter equation and the analytic structure of
the resultant amplitudes in the energy plane. This correlation is
studied for both the $\phi^2\sigma$ interaction Lagrangian and the
$\pi N$ system with $s$-, $u$-, and $t$-channel pole diagrams as
driving terms. We observe that the equal-time equation, which includes
some of the three-body unitarity cuts, gives the best agreement with
the Bethe-Salpeter result. This is followed by other 3-D
approximations that have less of the analytic structure.
\end{abstract}


\maketitle

\section{Introduction}\label{sec.1}

The interest in pion-nucleon ($\pi N$) scattering has shifted in recent years from
low energies to energies near and above the pion production
threshold.  This has been motivated by the growing importance of
the low energy baryon spectrum, and in particular the  $N^*(1440)$ and
$N^*(1535)$ in the study of QCD models. At these energies one needs to
make use of a relativistic formulation of $\pi N$ scattering, with the
Bethe-Salpeter (BS) equation~\cite{BS51} being a prime candidate. The
problem here is that the BS equation is four-dimensional, with a kernel
that has a complex analytic structure. This has inspired a number of
three-dimensional reductions to the BS equation that have been employed in
meson-exchange models of $\pi N$ and nucleon-nucleon ($NN$)
scattering. In fact, it has been established that one can
generate an infinity of such equations that satisfy two-body
unitarity~\cite{KL74}. The question that is often raised is which of
these equations is the optimum one to be used for a given problem.
To define such an optimum equation one needs criteria for the
definition of optimum.

One such criterion proposed by Gross~\cite{G82} stated that any
relativistic two-body equation should have the correct one-body
limit, i.e., in the limit as one of the masses goes to infinity
the equation reduces to either the Dirac or Klein-Gordon equation.
The Gross equation~\cite{G82} achieves this limit by placing one of
the particles on-mass-shell. This effectively reduces the BS equation
to a 3-D equation, and in the ladder approximation has the correct
one-body limit. The full BS equation does satisfy the one-body limit,
but the ladder BS equation fails to satisfy this condition. This could
be partly remedied in $NN$ scattering by including the crossed as well
as the ladder graphs in the kernel of the BS
equation~\cite{MW87,WM89,PW96}. For $\pi N$
scattering it is not clear as to how one may impose the one-body
limit short of resorting to the inclusion of the crossed-ladder
diagrams. Placing the pion on-shell~\cite{GS93} may be the optimal
solution for iterating the $u$-channel pole diagrams, but is not as
good for the $t$-channel pole diagrams~\cite{Pas98,PT99}.

An alternative scheme for developing  relativistic 3-D equations that
avoids the above ambiguities of the 3-D reductions, and is a natural
extension of the non-relativistic two-body scattering problem, is
based on a formulation of scattering theory using the generators of
the Poincar\'e group~\cite{BT53,KP91}. This formulation, which
introduces the interaction through one of the generators of the group,
satisfies two-body unitarity and can be extended to include three-body
unitarity. Recently, Fuda~\cite{F95,EF98} has adapted this approach to
the $\pi N$ system within the framework of the coupled channel
approach satisfying two-body unitarity only. To establish the
connection with meson exchange models~\cite{EF99}, the mass operator
$M$ in the interacting system is related to the mass operator in the
free system ($M_0$) by the linear relation $M=M_0+U$ in which $U$ is
the interaction. This interaction is defined on the Hilbert space of
baryon ($B$) and meson-baryon ($mB$) states, allowing for the coupling between
the two channels. To remove the energy dependence of the meson exchange
potentials they resort to Okuba's~\cite{O54} unitary transformation.
The source of ambiguities in this approach are: (i)~the choice for the
Poincar\'e generator in which the interaction is introduced, and
(ii)~the method by which the energy dependence is removed from the
interaction.

In the present analysis we propose to examine a third alternative to
the reduction of the BS equation from a four-dimensional to a
three-dimensional integral equation. Here we are motivated by trying
to preserve as much as possible the unitarity content of the BS
equation while preserving the two-body nature of the integral equation.
This is partly a result of the observation that the low
energy resonances observed in $\pi N$ scattering are near the
three-body threshold, and one may need more than just two-body
unitarity. It also could form the starting point of a three-body
formulation of $\pi N$ scattering without the complexity of the
three-particle BS equation.

To examine the importance of the role of unitarity in the reduction of
the BS equation, we examine a small class of three-dimensional
equations that preserve some of the unitarity cuts in the BS equation.
In particular, we consider the the Klein~\cite{K53} or ``equal-time''
(ET) equation~\cite{PW96,LA97}, the Cohen (C) approximation~\cite{C70}, the
instantaneous (I) approximation~\cite{S52,MW87,WM89,PT00} used recently by
Pascalutsa and Tjon for $\pi N$ scattering~\cite{PT00}, and the
Blankenbecler-Sugar (BbS) equation~\cite{BbS66}. Since there are an
infinite number of possible 3-D equations, there are many other equations
that we could compare to the BS equation. However, we have chosen to
consider only a small set of equations which differ in the analytic
structure of the resultant amplitude in the energy or
$\sqrt{s}$-plane.

In Sec.~\ref{sec.2} we first examine the analytic structure of the BS
amplitude for $\phi\phi$ scattering in the $\phi^2\sigma$ model by
performing a pinch analysis~\cite{PT66} on the relative energy
integration in the kernel of the integral equation. We will find in
this simple model that as we proceed from the BS equation to the ET,
Cohen, and finally to the instantaneous equation and the BbS equation,
we are including less and less of the analytic structure present in
the original BS amplitude. We then in Sec.~\ref{sec.3} proceed to present
the results of the same analysis for $\pi N$ scattering, within the
framework of $s$-, $u$- and $t$-channel exchange potentials. To
illustrate the magnitude and effect of these thresholds for both
$\phi\phi$ and $\pi N$ scattering, we present in Sec.~\ref{sec.4}
numerical results for both models over the region covering the
lowest three-body thresholds. Finally, in Sec.~\ref{sec.5} we give
some concluding remarks regarding the 3-D approximations to the BS
equation.

\section{Analytic Structure of the Scattering Amplitude}\label{sec.2}

Unlike two-body equations in three dimensions, the BS equation
generates unitarity thresholds by the pinching of relative energy
integration contour by the poles and
branch points of the integrand of the integral equation. To illustrate
this, we consider a simple model in which the interaction Lagrangian is
$g\phi^2\sigma$, where $g$ is the strength of the interaction, while
$\phi$ and $\sigma$ are the two scalar fields in the model. We will
further assume that the kernel of the BS equation for $\phi\phi$
scattering is approximated by one $\sigma$-exchange. The motivation for this
approximation is to simplify the analysis while maintaining the main
features of $NN$ scattering in terms of $t$-channel exchange, and
$\pi N$ scattering in terms of $s$-, $u$- and $t$-channel exchanges.

The BS equation for $\phi \phi$ scattering, after partial wave
expansion, is a two-dimensional integral equation of the form
\begin{equation}
T_\ell(\hat{q},\hat{q}^\prime;s) =
V_\ell(\hat{q},\hat{q}^\prime;s) - i
  \int _{-\infty}^\infty dq''_0\
  \int _{0}^\infty dq''
V_\ell(\hat{q},\hat{q}'';s)\,G(\hat{q}'',s)\,
T_\ell(\hat{q}'',\hat{q}^\prime;s)\ ,        \label{eq:1}
\end{equation}
where $\hat{q}=(q_0,q)$ is the zero component and magnitude of the
space component of the relative momentum, while
$P=k_1+k_2=(\sqrt{s},\textbf{0})$ is the total four-momentum in the
center of mass. Although there are several definitions for the
relative four-momentum~\cite{PJ91}, the most convenient, from the
point of view of the Wick rotation~\cite{Wi54}, is
$q=\frac{1}{2}(k_1-k_2)$. For scalar particles, the two-body
propagator in the center of mass is
\begin{equation}
G(\hat{q},s) =
\left[\left(\frac{1}{2}\sqrt{s}+q_0\right)^2 - E^2_q +
i\epsilon\right]^{-1}\
\left[\left(\frac{1}{2}\sqrt{s}-q_0\right)^2 - E^2_q +
i\epsilon\right]^{-1}\ ,                               \label{eq:2}
\end{equation}
where $E_q = \sqrt{q^2 + m^2}$, with $m$ the mass of the $\phi$ particle.
The partial wave potential due to the exchange of a $\sigma$ particle
of mass $\mu$ is given by
\begin{equation}
V_\ell(\hat{q},\hat{q}';s) = \frac{g^2}{(2\pi)^3}\,Q_\ell
\left(\frac{(q_0-q'_0)^2 -q^{2} -{q'}^{2} - \mu^{2}
+i\epsilon}{-2qq'}\right)\ .                          \label{eq:3}
\end{equation}
Here $Q_\ell$ is the Legendre function of the second kind, and for
$\ell=0$ this potential takes the form
\begin{equation}
V_0(\hat{q},\hat{q}';s) = \frac{g^2}{2(2\pi)^3}\ \log\left(
\frac{(q_0-q'_0)^2 - \omega^2_{q+q'}+i\epsilon}
{(q_0-q'_0)^2- \omega^2_{q-q'}+i\epsilon}\right)  \ , \label{eq:4}
\end{equation}
where $\omega_p=\sqrt{p^2+\mu^2}$. This form exhibits the logarithmic
branch points of the potential.
We observe that the potential is independent of the total
energy squared, $s$.

\begin{figure}[!]
     \includegraphics*[height=4.0 cm]{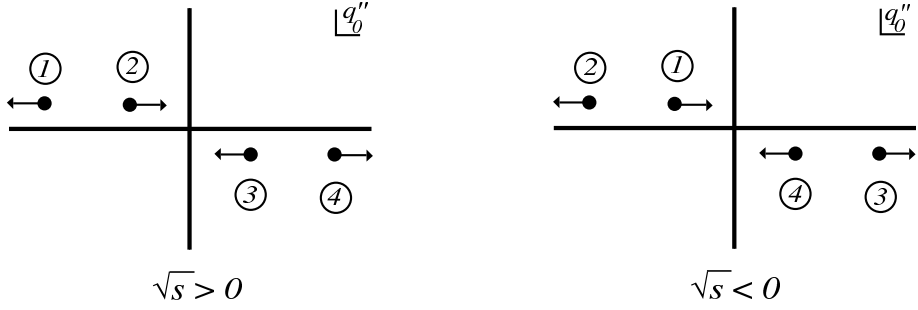}    
     \caption{The poles of the two-body propagator in the $q''_{0}$-plane.}
     \label{Fig.1}
\end{figure}

The analytic structure of the partial wave amplitude in the energy
($\sqrt{s}$) plane is determined by the pinching of the $q''_0$ integration
contour by the singularities of the integrand~\cite{E52,E62}. These
singularities arise from the potential $V_\ell(\hat{q},\hat{q}'';s)$,
the two-body propagator $G(\hat{q}'',s)$, and the scattering amplitude
$T_\ell(\hat{q}'',\hat{q}';s)$. The two-body propagator has four poles
in the $q''_0$-plane. These poles, labelled $1)\dots 4)$, are located at
\begin{eqnarray}
1)\quad q''_0 &=& -\frac{1}{2}\sqrt{s}-E_{q''}+i\epsilon\quad ;  \quad
2)\quad q''_0 =  \frac{1}{2}\sqrt{s}-E_{q''}+i\epsilon\ ; \nonumber \\
3)\quad q''_0 &=& -\frac{1}{2}\sqrt{s}+E_{q''}-i\epsilon\quad ; \quad
4)\quad q''_0 =  \frac{1}{2}\sqrt{s}+E_{q''}-i\epsilon\ ,   \label{eq:5}
\end{eqnarray}
and are illustrated in Fig.~\ref{Fig.1}.
As we increase $\sqrt{s}>0$ the poles $1)$ and $3)$ move to the left in the
$q''_0$-plane, while poles $2)$ and $4)$ move to the right.

To examine the conditions under which the $q''_{0}$ integration contour
gets pinched by the poles in the propagator, we consider
Fig.~\ref{Fig.1}. Here we have to ensure that the two poles that pinch
the contour of integration come from the Feynman propagators of two
different particles. As a result poles $1)$ and $4)$  (~or $2)$ and $3)$~)
can pinch the contour, while $1)$ and $3)$ (~or $2)$ and $4)$~) cannot
pinch the contour as they belong to the same particle propagator. This
results from the fact that while the Feynman propagator for a single
particle can be decomposed into a positive and a negative energy
component, these components cannot pinch the contour of
integration as they cannot generate any new thresholds. We
therefore have, for $\sqrt{s}>0$, the condition for the poles $2)$
and $3)$ to pinch the contour to be
\begin{equation}
     \frac{1}{2}\sqrt{s} - E_{q''} + i\epsilon =
     -\frac{1}{2}\sqrt{s} + E_{q''} - i\epsilon    \label{eq:6}
\end{equation}
or
\begin{equation}
     \sqrt{s} = 2\,E_{q''} - 2i\epsilon    \ .    \label{eq:7}
\end{equation}
Here we note that if we deform the $q''$ contour of integration in
Eq.~(\ref{eq:1}), e.g. $q''\rightarrow q''\,e^{-i\phi}$,
then $E_{q''}$ will acquire a negative imaginary part and as a result
the $q_{0}''$ integration contour will not get pinched. This
is true for all values of $q''$ except $q''=0$. Therefore we may
conclude that the pinching of the $q''_{0}$ contour generates a branch
point at $\sqrt{s}=2m$, which is the threshold for two-body scattering.

The same analysis can be carried through for $\sqrt{s}<0$. In this
case the contour of integration in the $q''_{0}$ variable is pinched by the
poles 1) and 4), with the condition for the pinch being
\begin{equation}
     \sqrt{s} = -2\,E_{q''} + 2i\epsilon   \ . \label{eq:8}
\end{equation}
This will generate a branch point at $\sqrt{s}=-2m$. This result
establishes the fact that if we take both branch points resulting
from the pinching of the integration contour by the poles of the
two-body propagator, we get an amplitude
that is analytic in $s$. In this way charge conjugation symmetry is
preserved~\cite{PT98}. However, if we only include the branch point at
$\sqrt{s}=2m$, then we have only a right hand cut in the
$\sqrt{s}$-plane and the resultant amplitude is a function of
$\sqrt{s}$, i.e., it satisfies two-body unitarity, but
violates charge conjugation symmetry.

\begin{figure}[!]
    \includegraphics*[height=4.0 cm]{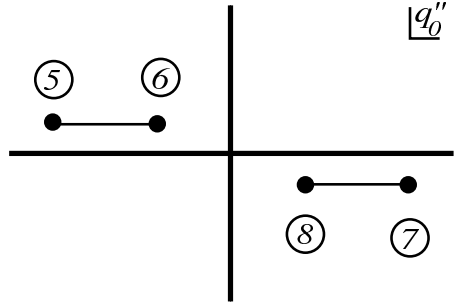}
     \caption{The branch points of the potential in the
     $q''_{0}$-plane.}
     \label{Fig.2}
\end{figure}

The potential $V_\ell(\hat{q},\hat{q}'';s)$ 
gives rise to two pairs of logarithmic
branch points in the $q''_0$-plane. These branch points, labelled $5)
\dots 8)$, are located at
\begin{eqnarray}
5)\quad q''_0 = q_0 - \omega_{q+q''} + i\epsilon\quad ;&\quad
6)\quad q''_0 = q_0 - \omega_{q-q''} + i\epsilon\ ;\nonumber \\
7)\quad q''_0 = q_0 + \omega_{q+q''} - i\epsilon\quad ;&\quad
8)\quad q''_0 = q_0 + \omega_{q-q''} - i\epsilon \ ,   \label{eq:9}
\end{eqnarray}
with branch points $5)$ and $6)$ ( $7)$ and $8)$ ) being connected by
a logarithmic branch cut. These are illustrated in
Fig.~\ref{Fig.2}. Since the potential is due to the exchange of a
scalar particle, the potential, before partial wave expansion, is
proportional to the Feynman propagator for that scalar. Here again the
decomposition of the propagator into positive and negative energy
components gives rise to the two logarithmic cuts in Fig.~\ref{Fig.2},
and each will  give rise to a separate term in the multiple scattering
series.

We now can examine the pinching of the contour of integration in the
$q''_{0}$-plane by a combination of a propagator pole and a potential
branch point. Take for example the pole 2) with the branch point 8). The
condition for the pinch is
\begin{equation}
     \frac{1}{2}\sqrt{s} - E_{q''} + i\epsilon =
     q_{0} + \omega_{q-q''} - i\epsilon                \label{eq:10}
\end{equation}
or
\begin{equation}
     \frac{1}{2}\sqrt{s} = q_{0} + E_{q''} +\omega_{q-q''} -
     2i\epsilon \equiv p_{0}\ .                        \label{eq:11}
\end{equation}
Here again contour deformation, e.g. $q''\rightarrow
q''e^{-i\phi}$, will stop us from pinching the contour since both
$E_{q''}$ and $\omega_{q-q''}$ acquire a finite imaginary part. The
exception again is the point $q''=0$. In this case the value of $q''$
for which we get the pinching of the contour depends on $\hat{q}$. To
determine the position of the branch point in the $\sqrt{s}$-plane, we
need to find the minimum value of $p_{0}$ or $\sqrt{s}$ with respect
to $q''$ for which we have a pinch, i.e.,
\[
\frac{\partial p_{0}}{\partial q''} = \frac{q''}{E_{q''}} -
\frac{(q-q'')}{\omega_{q-q''}} = 0\ .
\]
This equation is used to determine $q''$ in terms of $q$, and gives
\[
q'' = \frac{m}{m+\mu}\,q\ .
\]
We now can establish that the off-mass-shell scattering amplitude has
a branch cut in the $q_{0}$-plane at~\cite{PT66}
\begin{equation}
     q_{0} = \frac{1}{2}\sqrt{s} - \sqrt{q^{2} + (m+\mu)^{2}}
           + 2i\epsilon\ .                           \label{eq:12}
\end{equation}
This gives the singularity in the relative energy $q''_{0}$ for the
off-mass-shell amplitude that is being integrated over in
Eq.~(\ref{eq:1}).

\begin{figure}[!]
    \includegraphics*[height=4.0 cm]{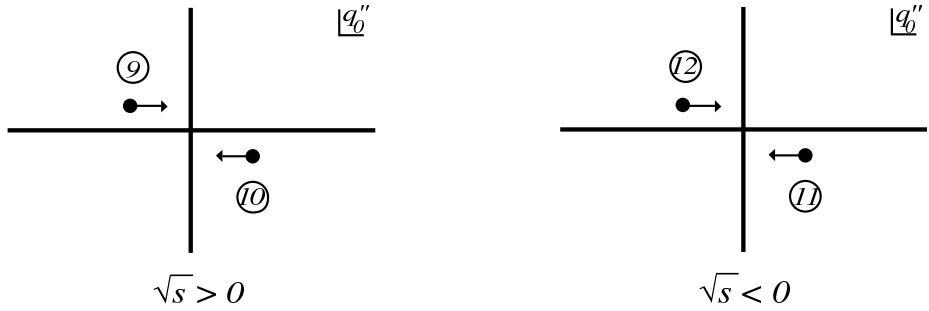}
\caption{The branch points of the amplitude resulting from the
     pinching of the integration contour by the propagator poles and potential
     branch points in the $q''_{0}$-plane. The arrows indicate the
direction in which the branch points move with increasing $s$.}
     \label{Fig.3}
\end{figure}

Following the same procedure, we examine the pinching of the contour
by the singularities $3)$ and $6)$  for $\sqrt{s}>0$, while for
$\sqrt{s}<0$ the contour is pinched by either $1)$ and $8)$, or
$4)$ and $6)$. As a result the off-mass-shell amplitude in the
integral has four possible branch cuts in the $q''_{0}$-plane.
These are at
\begin{equation}
9)\quad q''_{0} = \frac{1}{2}\sqrt{s} - {\cal E}_{q''}
               + 2i\epsilon   \  ; \quad
10)\quad q''_{0} = - \frac{1}{2}\sqrt{s} + {\cal E}_{q''}
               - 2i\epsilon                       \label{eq:13}
\end{equation}
for $\sqrt{s}>0$, and for $\sqrt{s}<0$ the branch cuts are given by
\begin{equation}
11)\quad q''_{0} = \frac{1}{2}\sqrt{s} + {\cal E}_{q''}
               - 2i\epsilon   \  ; \quad
12)\quad q''_{0} = - \frac{1}{2}\sqrt{s} - {\cal E}_{q''}
               + 2i\epsilon\ ,                   \label{eq:14}
\end{equation}
where ${\cal E}_{q}=\sqrt{q^{2}+(m+\mu)^{2}}$. In Fig.~\ref{Fig.3} we
illustrate the positions of the corresponding branch points in the
$q''_{0}$-plane. At this stage we should recall that the singularities
$9)$ and $10)$ are in different terms in the multiple scattering
series, and therefore cannot pinch the contour of integration as may
be inferred from Fig.~\ref{Fig.3}.

Since the singularities $9)\dots 12)$ are in the off-mass-shell BS
amplitude, they are present in the integrand of the integral equation,
Eq.~(\ref{eq:1}). These singularities in the amplitude can, in
conjunction with the propagator poles or potential branch points,
pinch the $q''_{0}$ integration contour and generate further singularities
in the amplitude on the left hand side of Eq.~(\ref{eq:1}).

Let us first consider the pinching of the contour by the amplitude
singularities and the propagator poles. For $\sqrt{s}>0$ we now can
have the integration contour pinched by the singularities $2)$ and
$10)$ (~or $3)$ and $9)$~). The condition for the pinching between the
pole $2)$ and the singularity $10)$ is
\begin{equation}
     \frac{1}{2}\sqrt{s} - E_{q''} +i\epsilon =
     -\frac{1}{2}\sqrt{s} + {\cal E}_{q''} - 2i\epsilon \label{eq:15}
\end{equation}
or
\begin{equation}
     \sqrt{s} = E_{q''} +{\cal E}_{q''} - 3i\epsilon\ .\label{eq:16}
\end{equation}
This pinching of the contour generates a singularity for $q''=0$
at $\sqrt{s} = 2m+ \mu $. This is the three-body threshold for
$\sigma$ production. We can proceed in the same manner to examine the
other possible pinching of the contour in the $q''_{0}$-plane by the
poles of the propagator and the singularities of the off-mass-shell
amplitude. These will generate thresholds in the $\sqrt{s}$-plane at
\begin{equation}
     \sqrt{s}=\pm (2m+\mu)                 \ .        \label{eq:17}
\end{equation}
This gives us one of the thresholds needed for the equation to
satisfy three-body unitarity. At this stage our amplitude is a
function of $s$ and therefore satisfies the requirement of charge
conjugation symmetry~\cite{PT00}.

We can also pinch the $q''_{0}$ integration contour by a
singularity of the amplitude and a branch point of the potential. For
example, if we pinch the contour of integration with singularities
$6)$ and $10)$, we get a branch cut in the off-shell amplitude at
$q_{0} =\frac{1}{2}\sqrt{s} + \sqrt{q^{2} + (m+2\mu)^{2}} -
3i\epsilon$. This in turn gives rise to a branch point in the
off-mass-shell amplitude in the integral part of Eq.~(\ref{eq:1}),
and with the help of the propagator pole $2)$, will pinch the
integration contour to generate the four-body threshold at
$\sqrt{s} = 2m+2 \mu$.

One can continue to include higher singularities into the off-mass-shell
amplitude by considering the pinching of the contour of integration by
singularities of the amplitude and those of the potential. These
off-mass-shell singularities when included in the integral part of
Eq.~(\ref{eq:1}) can, with the help of the propagator pole, pinch the
contour in  $q''_{0}$, and generate further thresholds in the
$\sqrt{s}$-plane. These thresholds are for two-, three-, ...
$\sigma$ production and are located at
\begin{equation}
s = (2m+n\mu)^2\quad\mbox{for}\quad n=1, 2, \dots\ .\label{eq:18}
\end{equation}
In this way we have demonstrated that the BS amplitude has not only
the two-body threshold, but all $n$-body thresholds. Unfortunately, in
the ladder approximation with undressed propagators and vertices, the
Bethe-Salpeter amplitude does not give the full contribution to
$n$-body unitarity.

We now turn to three-dimensional reductions of the BS equation
(also known as quasi-potential equations), and
illustrate how they have a smaller subset of the thresholds and/or
off-mass-shell singularities than  the BS amplitude. As a first
approximation we consider the equal-time equation. This was originally
suggested by Klein~\cite{K53} and has more recently been revisited for
the $\phi^2\sigma$ model~\cite{LA97} and the nucleon-nucleon
interaction~\cite{PW96}, where the problem has been restated in a more
formal manner which allows one to consistently improve upon the 3-D
results to ultimately achieve the BS result. In this case the 3-D
equation is given by~\cite{LA97}
\begin{equation}
T_1 = K_1 + K_1\bra\,G\,\ket T_1    \ ,         \label{eq:19}
\end{equation}
where the 3-D amplitude $T_1$ and potential $K_1$ are defined in terms
of integrals over the relative energy $q_{0}$, i.e.,
\begin{eqnarray}
T_1 &=& \bra\,G\,\ket^{-1}\bra\,G\,T\,G\,\ket\
\bra\,G\,\ket^{-1} \nonumber \\
K_1 &=& \bra\,G\,\ket^{-1}\bra\,G\,V\,G\,\ket\
\bra\,G\,\ket^{-1} \ ,                          \label{eq:20}
\end{eqnarray}
where the brackets in the above expressions are defined as
\begin{equation}
\bra\,A\,\ket = \int dq_0\,dq'_0\ A(q_0,\textbf{q};q'_0,\textbf{q}';s)
\ .                                               \label{eq:21}
\end{equation}
In this case the thresholds are generated by the double integral over
the relative energies $q_0$ and $q'_0$ in the initial and final state in
$\bra\,G\,V\,G\,\ket$. Here, (see Appendix~\ref{app.1})
for $\sqrt{s}>0$ the $q'_0$ integration will generate:
(i)~the elastic threshold at $\sqrt{s}=\pm 2m$ by the pinching of the two
poles of the propagator to the right of $V$ in Eq.~(\ref{eq:20});
(ii)~the pinching of the propagator poles with the potential branch
points will generate the off-mass-shell singularities $9)$ and $10)$
in the $q_0$-plane for $\sqrt{s}>0$. For $\sqrt{s}<0$, the
singularities $11)$ and $12)$ are generated in the $q_0$-plane.
These singularities, with the help of the poles from the propagator
on the left of $V$ in Eq.~(\ref{eq:20}), will generate branch points
at $\sqrt{s}=\pm(2m+\mu)$ (see Appendix~\ref{app.1}).

Thus the ET equation includes, in addition to two-particle
unitarity, the contribution to three-body unitarity from the
$\sigma$-exchange diagram. Unfortunately, as was the case with the BS
equation, these thresholds are only part of the three-body
unitarity cuts that should be present. Additional contributions to
three-body unitarity come from the dressing of the $\phi$
propagators, the $\phi^2\sigma$ vertex and from higher-order
contributions to the potential such as the crossed box diagram.

In the Cohen~\cite{C70} approximation it is assumed that the amplitude
in Eq.~(\ref{eq:1}) does not depend on the relative energy $q''_0$. As a
result, we can perform the $q''_0$ integration, i.e.,
\begin{equation}
\int _{-\infty}^{\infty}dq''_0\ V_\ell(\hat{q},\hat{q}'';s)\
G(\hat{q}'',s)                    \ .               \label{eq:22}
\end{equation}
In this case the only pinching of the $q''_0$ integration contour is:
(i)~between two propagator poles, and (ii)~between a propagator pole
and a potential branch point. Here, in addition to the two-body
threshold at $\sqrt{s}=\pm 2m$ generated by the pinching of the
contour by the propagator poles, we have singularities
in the off-mass-shell amplitude resulting from the pinching of the
propagator poles and the potential branch points. These correspond to
the singularities $9)$ and $10)$, $11)$ and $12)$. This introduces an
inconsistency in the equation to the extent that
the analytic properties of the amplitude as predicted by the integral
part of the integral equation are not consistent with those assumed in
the reduction of the dimensionality of the equation from four to
three. To resolve this inconsistency it is assumed that the relative
energy in the amplitude takes on its on-mass-shell value,
i.e., $q_0=0$. In that case the amplitude has, in addition to
the two-body threshold, a branch cut at $\sqrt{s}=2{\cal E}_{q}\pm
4i\epsilon$ and as a result a threshold in $\sqrt{s}$, at $\sqrt{s}=
\pm 2(m+\mu)$. This indicates that the Cohen amplitude has one of the
four-particle thresholds, but no three-body thresholds.

In the instantaneous approximation, the
inconsistency has been removed by assuming a static potential and
therefore fixing the value of the relative energy in the
potential~\cite{PT00} (e.g. $q_{0}=q''_0=0$). As a result
the only singularity generated by the pinching of the $q''_0$ contour
is by the poles of the two-body propagator. These generate the
thresholds at $\sqrt{s}=\pm 2m$. 
We should point out that for $\phi\phi$ scattering the Blankenbecler-Sugar
equation is identical to the instantaneous approximation.

\section{Analytic structure of the $\pi N$ amplitude}\label{sec.3}

We now turn to the analytic structure of the $\pi N$ amplitude. Here
we have three classes of diagrams that can contribute to the potential
at the one particle exchange level: the $s$-, $u$-, and
$t$-channel pole diagrams. The $s$-channel pole diagrams do not
contribute any singularities to the integral part of the BS equation,
Eq.~(\ref{eq:1}). As a result, the only thresholds generated by the 
$s$-channel pole diagrams in the potential are those resulting from
the pinching of the integration contour by the $\pi N$ propagator and amplitude
singularities. These singularities are also present when the potential
is due to $u$- and $t$-channel pole diagrams, and need not be
considered until we examine the $u$- and $t$-channel pole diagrams.

For the $u$- and $t$-channel pole diagrams, the potential takes the
form
\begin{equation}
V(q,q';P) = \frac{ F(q,q';P)}
                  {(q+\zeta q')^2 - m_H^2 + i\epsilon}\ ,\label{eq:23}
\end{equation}
where $\zeta=1$ for the $u$-channel pole diagrams, and $-1$ for
the $t$-channel pole diagrams. Here $m_H$ is the mass of the exchanged
particle. Thus for $u$-channel pole diagrams $m_H=m_N,m_\Delta,\cdots$,
while for $t$-channel diagrams $m_H\rightarrow \mu_{H} = \mu_\rho,
\mu_\sigma,\cdots$. The
function $F(q,q';P)$ depends on the spin of the exchanged
hadron and the form of the coupling at the two vertices in the
diagrams. Since the function $F$ has no singularities that could take
part in the pinching of the integration contour, we could assume
$F=1$ if we are willing to neglect form factors associated with the
vertices, and we do not consider the values of the discontinuities across
any cuts in the amplitude. This is equivalent to assuming all particles
are scalars, and all physical thresholds are independent of the spin and
isospin of the particles involved. Consequently, the BS equation
maintains the form given in Eq.~(\ref{eq:1}) with the contribution to
the partial wave potential for $u$- and $t$-channel pole diagrams given
for $\ell=0$ by
\begin{equation}
V_0(\hat{q},\hat{q}') = \zeta\frac{g^2}{2(2 \pi)^3}\log\left\{
\frac{(q_0+\zeta q_0')^2-(q+\zeta q')^2 - m_H^2 + i\epsilon}
{(q_0+\zeta q_0')^2-(q-\zeta q')^2 - m_H^2 + i\epsilon}\right\}
                                                     \ .\label{eq:24}
\end{equation}

The $\pi N$ propagator can be divided into two contributions,
one from the positive energy component of the nucleon propagator,
and the second from the negative energy component. We have in the
center of mass
\begin{equation}
G_{\pi N}(q_0,\textbf{q};s) = G^{\bar{u}u}_{\pi N}(\hat{q},s)\Lambda^+(\textbf{q}) -
G^{\bar{v}v}_{\pi N}(\hat{q},s)\Lambda^-(-\textbf{q}) \ ,    \label{eq:25}
\end{equation}
where the energy projection operators are written in terms of Dirac spinors as
\begin{subequations}
\begin{eqnarray}
\Lambda^+(\textbf{q}) & = & \sum _r u_r(\textbf{q}) 
\bar u_r(\textbf{q}) \ ,                            \\
\Lambda^-(\textbf{q}) & = & -\sum _r v_r(\textbf{q}) 
\bar v_r(\textbf{q}) \ .                            
\end{eqnarray}
\end{subequations}
Also
\begin{subequations}
\label{eq:bsegf}
\begin{eqnarray}
G^{\bar{u}u}_{\pi N}(\hat{q},s)&=& \frac{m_N}{E_q}
\frac{1}{\alpha _N \sqrt{s}+q_0 - E_q+i\epsilon}\,
\frac{1}{(\alpha _{\pi} \sqrt{s}-q_0)^2 - \omega^2_q+i\epsilon} \ , \label{eq:28} \\
G^{\bar{v}v}_{\pi N}(\hat{q},s)&=& \frac{m_N}{E_q}
\frac{1}{\alpha _N \sqrt{s}+q_0 + E_q-i\epsilon}\,
\frac{1}{(\alpha _{\pi} \sqrt{s}-q_0)^2 -\omega^2_q+i\epsilon}\ ,\label{eq:29}
\end{eqnarray}
\end{subequations}
with $E_q=\sqrt{\textbf{q}^2+m^2_N}$ and $\omega_q=\sqrt{\textbf{q}^2
+ \mu_{\pi}^2}$. In addition $\alpha _N$ and $\alpha _{\pi}$ are functions of 
$s$ such that $\alpha _N + \alpha _{\pi} = 1$. The solution of the BS equation
does not depend on the choice of $\alpha _N(s)$ and $\alpha _{\pi}(s)$, so for
convenience, in our discussion of thresholds generated by the BS equation,
we use the simplest possibility: $\alpha _N = \alpha _{\pi} = 1/2$.

We note that for $G_{\pi N}$ the number of poles is still four, and
the pinching of the integration contour by the propagator poles
generate the elastic thresholds at $\sqrt{s}=\pm(m_N+\mu_{\pi})$.
Therefore it is only the pinching between the propagator poles and the
potential branch points that needs to be considered.

\subsection{$\mathbf{t}$-channel pole diagrams}\label{sec.3.a}

We first examine the thresholds generated by the $t$-channel pole
diagrams. These are the result of pinching of the contour of
integration in the $q_0''$ plane by the poles of the propagator
and the branch points of the potential. Here we follow the same
procedure adopted in the last section for $\phi\phi$ scattering in
the $\phi^2\sigma$ model, and state the results. For $\sqrt{s}>0$,
the pinching of the potential singularity with the propagator pole
generates two branch points in the off-mass-shell amplitude at the
relative energy
\begin{equation}
    q_{0} =\mp\frac{1}{2}\sqrt{s} \pm \sqrt{q^{2}+(m_{N}+\mu_{H})^{2}}
             \mp2\,i\,\epsilon          \ .    \label{eq:30}
\end{equation}
These two branch points are also present in the off-mass-shell
amplitude in the integral part of the integral equation
Eq.~(\ref{eq:1}), and with the help of either the propagator poles or
the potential branch point, can pinch the integration contour to
generate further branch points. In conjunction with the propagator
poles the branch points in Eq.~(\ref{eq:30}) generate the threshold at
\begin{equation}
     \sqrt{s} = (m_{N}+\mu_{\pi}+\mu_{H})       \ . \label{eq:31}
\end{equation}
These are thresholds for $\rho$ and $\sigma$ production in a model in
which the $t$-channel poles are represented by $\rho$ and $\sigma$
exchanges.

The pinching of the amplitude singularities in Eq.~(\ref{eq:30}) and
the potential branch points will, for $\sqrt{s}>0$, generate branch
points in the off-mass-shell amplitude 
\begin{equation}
     q_{0}= \mp\frac{1}{2}\sqrt{s}\pm\sqrt{q^{2}+(m_{N}+\mu_{H}
     +\mu_{H'})^{2}} \mp 2\,i\,\epsilon\ .      \label{eq:32}
\end{equation}
Here, as in the case of scalar particles considered in
Sec.~\ref{sec.2}, the Bethe-Salpeter equation has all the thresholds for the
production of the mesons present as $t$-channel pole diagrams in the
potential.  Thus the amplitude branch point in Eq.~(\ref{eq:32})
with the help of the propagator pole will generate the four-body
threshold at
\begin{equation}
     \sqrt{s} = (m_{N}+\mu_{\pi}+\mu_{H}+\mu_{H'})\ ,\label{eq:33}
\end{equation}
which is present in the off-mass-shell box diagrams given by
$G\,V_{t}\,G\,V_{t}\,G$.

\subsection{$\mathbf{u}$-channel pole diagrams}\label{sec:3.b}

Turning to the $u$-channel pole diagrams, the branch points of the
potential with the help of the propagator poles generate two
off-mass-shell branch points at
\begin{equation}
q_{0} = \mp\frac{1}{2}\sqrt{s} \pm \sqrt{q^{2} + (m_{N}+m_{H})^{2}}
         \mp 2\,i\,\epsilon                 \ .\label{eq:34}
\end{equation}
These amplitude branch points in conjunction with the propagator pole
generate three-body thresholds at
\begin{equation}
     \sqrt{s} = 2m_{N}+m_{H}\quad\mbox{and}\quad \sqrt{s} =2\mu_{\pi} +
     m_{H}       \ .                      \label{eq:35}
\end{equation}
To generate the four-body threshold we have to examine the three
classes of off-mass-shell box diagrams corresponding to
$G\,V_{u}\,G\,V_{u}\,G$, $G\,V_{t}\,G\,V_{u}\,G$ and
$G\,V_{u}\,G\,V_{t}\,G$. The first class gives rise to the threshold at
\begin{equation}
     \sqrt{s} = (m_{N} + \mu_{\pi} + m_{H} + m_{H'})\ .\label{eq:36}
\end{equation}
This corresponds to first pinching the integration contour by the
singularities in Eq.~(\ref{eq:34}) and the branch point of the
$u$-channel pole potential. This is to be followed by the pinching of
the integration contour by the resultant branch point and the
propagator pole. The second class of box diagrams
$G\,V_{t}\,G\,V_{u}\,G$ will generate thresholds at
\begin{equation}
     \sqrt{s} = (2\,m_{N}+ m_{H}+\mu_{H'}) \quad\mbox{and}\quad
     \sqrt{s} = (2\,\mu_{\pi}+m_{H}+\mu_{H'})\ .  \label{eq:37}
\end{equation}
Finally the third class of box diagram $G\,V_{u}\,G\,V_{t}\,G$ gives
four-body thresholds at
\begin{equation}
     \sqrt{s} = (2m_{N}+m_{H}+\mu_{H'})\quad\mbox{and}\quad
     \sqrt{s} = (2\mu_{\pi}+ m_{H}+\mu_{H'})\ ,   \label{eq:38}
\end{equation}
which are identical to those in $G\,V_{t}\,G\,V_{u}\,G$ as expected.
In the above $m_{H}=m_{N}, m_{\Delta},\cdots$ and $\mu_{H} =
\mu_{\rho}, \mu_{\sigma},\cdots$. We could continue this procedure to
generate all the higher order thresholds in the Bethe-Salpeter
equations as we did in Sec.~\ref{sec.2}.

The above analysis is for $\sqrt{s}>0$, and similarly we could generate
the mirror thresholds by examining the case $\sqrt{s}<0$. In this way
we establish that the amplitude is a function of $s$ rather than
$\sqrt{s}$.

The point to be emphasised in establishing the positions of the
thresholds in the $s$-plane is that these thresholds, included in the
ladder BS equation, are not sufficient for the amplitude to satisfy
three-, four- and higher-body unitarity. This means that we have some
three-body thresholds, but we don't have three-body unitarity. Only
two-body unitarity is included completely. To
include three-body unitarity we would need to carry through a Faddeev
decomposition of the amplitude, which would result in a set of coupled
equations that are significantly more complex to solve~\cite{PA96}.

\subsection{Thresholds in the 3-D equations}\label{sec:3.c}

We now turn to the three-dimensional reduction of the BS equation
for the $\pi N$ system. We consider the equal-time, Cohen,
instantaneous, and the Blankenbecler-Sugar 
equations, each of which has different unitarity structure in the
$s$-plane. The BbS equation is just one example of an infinite number
of 3-D equations that are derived from the BS equation by replacing
the $\pi N$ propagator $G_{\pi N}$ with a new propagator containing
a $\delta$-function which fixes the relative energy. As a result, the only
threshold included is the two-body unitarity cut. For more information
about the 3-D equations, see Appendix~\ref{app.2}. 

The instantaneous and BbS equations generate thresholds at
$\sqrt{s}=\pm(m_{N}+\mu_{\pi})$, however, the coupling to negative
energy states is neglected in the BbS equation.
The equal-time
or Klein equation has, in addition to the two-body thresholds at
$\sqrt{s}=\pm(m_{N}+\mu_{\pi})$, the three-body
thresholds as presented above for the ladder BS equation. This includes
\begin{equation}
     \sqrt{s} = m_{N}+\mu_{\pi}+\mu_{H}
     \ ,\quad\sqrt{s} =  2m_{N}+m_{H}\ ,
      \quad\mbox{and}\quad \sqrt{s} = 2\mu_{\pi}+ m_{H}\ .\label{eq:39}
\end{equation}
The Cohen approximation, in addition to the two-body
thresholds at $\sqrt{s} = \pm(m_{N}+\mu_{\pi})$, has 
thresholds resulting from the pinching of the contour of integration
by the propagator poles and the potential branch points with $q_{0}$
on-mass-shell, i.e., $q_{0}=(m_N^2-\mu_{\pi}^2)/(2 \sqrt{s})$. For the 
$t$-channel pole diagram,
we get for $\sqrt{s}>0$ the branch point at
\begin{equation}
     \sqrt{s} = m_{N}+\mu_{\pi}+ 2\mu_{H}\  .   \label{eq:40}
\end{equation}
On the other hand, the $u$-channel pole diagram gives thresholds at
\begin{equation}
     \sqrt{s} = -\mu_{\pi}+3m_{N}+2m_{H}\quad\mbox{and}\quad
     \sqrt{s} = -m_{N}+3\mu_{\pi}+2m_{H}\ .    \label{eq:41}
\end{equation}
These are basically the four-body thresholds the Cohen approximation had
in the scalar case considered in Sec.~\ref{sec.2}. Here again the
analysis is for $\sqrt{s}>0$, and can be extended to get the mirror
thresholds for $\sqrt{s}<0$.

\section{Numerical Results}\label{sec.4}

To examine the importance of the higher unitarity cuts as we proceed
from the BS equation to the various 3-D reductions, we will present numerical
results for the two models ($\phi^2\sigma$ and $\pi N$) under
consideration. Since thresholds do not depend on the spin and
isospin of the fields involved, we have chosen the $\phi^2\sigma$
model as an approximation to the nucleon-nucleon interaction with one
pion exchange in the deuteron channel. There are no
resonances in the system, and we expect the higher unitarity cuts to be 
of no great importance as is the case for the $^3 S_1$
$NN$ channel. On the other hand, for the $\pi N$ system the 
dominance of the $\Delta(1232)$ will allow us to examine the effect of 
the higher unitarity cuts on the width and position of a resonance.

\begin{figure}[!]
     \centering\includegraphics[scale=0.7]{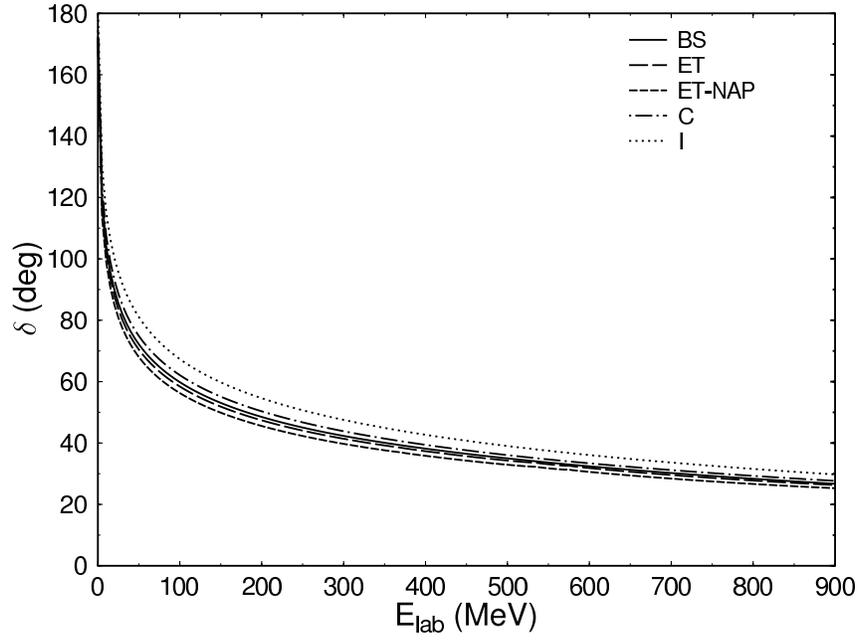}
\caption{The $S$-wave phase shifts for $\phi\phi$ scattering with one
$\sigma$-exchange as the potential using the BS, ET, ET-NAP, C, and I 
equations.}\label{Fig.4} 
\end{figure}

To emulate the $NN$ system in the $\phi^2\sigma$ model with $\sigma$
exchange as the potential, we have chosen the mass of the $\phi$ ($N)$ 
to be 1~GeV, while the $\sigma$ ($\pi$) has a mass of $0.15$~GeV. For
$S$-wave scattering we have adjusted the coupling constant so that the 
$\phi\phi$ system has a bound state with a binding energy $2.2$~MeV
corresponding to the deuteron. In Fig.~\ref{Fig.4} we present the phase 
shifts resulting from the solution of the BS equation as well as the
equal-time, Cohen, and the instantaneous 3-D equations. Also 
included are the
phase shifts resulting from the equal-time equation after excluding
the negative energy component of the propagator (ET-NAP),
i.e., we have excluded the left hand cut and hence violated charge
conjugation symmetry. This situation corresponds to time-ordered
perturbation theory~\cite{LA97}, which has been used frequently in models of
$NN$ scattering (see, e.g., Ref.~\cite{MC89})
and $\pi N$ scattering (see, e.g., Ref.~\cite{JL94}).
From the results in Fig.~\ref{Fig.4} we find, as expected,
no great variation as one excludes higher unitarity cuts. On the other 
hand, the differences are consistent with the unitarity analysis of the 
last section in that the best approximation is the ET equation,
followed by C, ET-NAP, and finally I. It is 
interesting to note that the inclusion of the three-body unitarity cut in the ET-NAP 
equation is more important than satisfying charge conjugation which 
is included in the I equation.

\begin{figure}[!]
    \centering\includegraphics[scale=0.7]{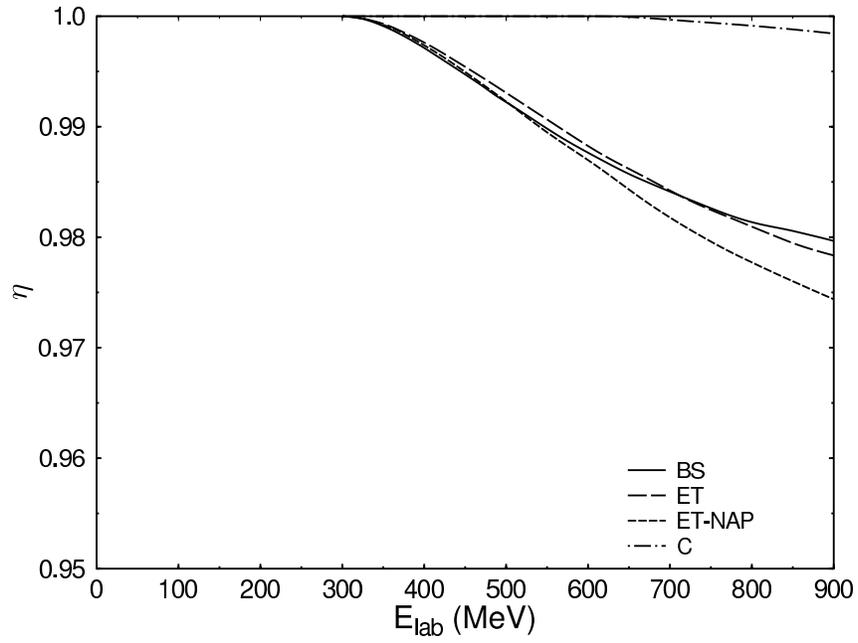}
\caption{The inelasticity $\eta$ for the BS, ET, ET-NAP, and C
equations.}\label{Fig.5} 
\end{figure}

A better test of the role of higher unitarity cuts is to examine the
inelasticity for the different approximations to the BS equation, the
results of which are shown in Fig.~\ref{Fig.5}. In
this case results for the I equation have 
not been included since the inelasticity is always unity, as this 
equations include the two-body unitarity cuts only. The agreement 
between the BS and ET equations is a result of the fact that both  
equations have the same three-body unitarity cuts. The Cohen 
equation does not have a three-body unitarity cut, and as a result, the 
inelasticity for this equation is less than one only for energies above the 
four-body unitarity threshold. At this stage the results for the BS 
and ET equations start to deviate as the latter does not include the 
four-body thresholds. Although we can observe differences between the 
equations in the inelasticity, in all cases the inelasticity is very 
small for this problem.

From the above analysis we may conclude that for energies close to
the $\sigma$ production threshold, the ET equation handles the
inelasticity much better than any of the other 3-D
equations as it includes the same three-body unitarity cuts that the
BS equation has.

We now turn to the $\pi N$ system, and consider a model
in which the potential includes $N$ and $\Delta$ 
$s$- and $u$-channel pole diagrams as well as $\sigma$ and $\rho$
$t$-channel pole diagrams. To solve the BS equation or one of the
3-D equations for the $\pi N$ system we need
to introduce some form of regularization. This has been achieved by
introducing two types~\cite{LA99} of form factors for each of the 
vertices in the diagrams that constitute the potential. 
\begin{itemize} 
\item Type I: Since the vertices in the
potentials have three legs, we can write a form factor for each vertex 
as the product of three functions of the four-momenta of the three 
legs, i.e.,
\begin{equation}
f_{\alpha\beta\gamma}(p_\alpha^2,p_\beta^2,p_\gamma^2) =
f_\alpha(p_\alpha^2)\,f_\beta(p_\beta^2)\,
f_\gamma(p_\gamma^2)                         \ ,\label{eq:42}
\end{equation}
where $\alpha$, $\beta$, and $\gamma$ refer to the three legs of the
vertex. The function $f_\alpha(p_\alpha^2)$ is taken to be 
\begin{equation}
f_\alpha(p_\alpha^2) = \left(\frac{\Lambda_\alpha^2 - m_\alpha^2}
{\Lambda_\alpha^2 - p_\alpha^2}\right)^{n_\alpha}\ ,\label{eq:43}
\end{equation}
with $n_\alpha=1$.
Here $m_\alpha$ is the mass of the particle, while $\Lambda_\alpha$
is the cutoff mass that is a parameter of the potential. Since such
form factors have poles, they can generate unphysical thresholds. To
minimize the effect of these unphysical thresholds, we have
constrained the cutoff masses $\Lambda_\alpha$ to ensure 
that these thresholds are at much higher energies than any of the
physical thresholds we are examining.
\item Type II: To reduce the number of cutoff masses, and at the same 
time employ a form factor that is more commonly used, we have taken
the form factor to depend on the four-momentum of the pion only,
i.e.,
\begin{equation}
f_{\alpha\beta\gamma}(p_\alpha^2,p_\beta^2,p_\gamma^2) =
f_\pi(p_\pi^2)\ .                                 \label{eq:44}
\end{equation}
Here we present results for $n=4$ only.
\end{itemize}
The parameters of the BS potential and form
factors have been adjusted to fit the $\pi N$ data up to a laboratory
energy of 360~MeV~\cite{LA99}. We use the BS parameters in the different
3-D equations, i.e., we do not refit the parameters using each of the
scattering equations. This is so that the effects of the different 3-D reductions
can be easily compared to the BS equation.

\begin{figure}[!]
   \centering\includegraphics[scale=1.0]{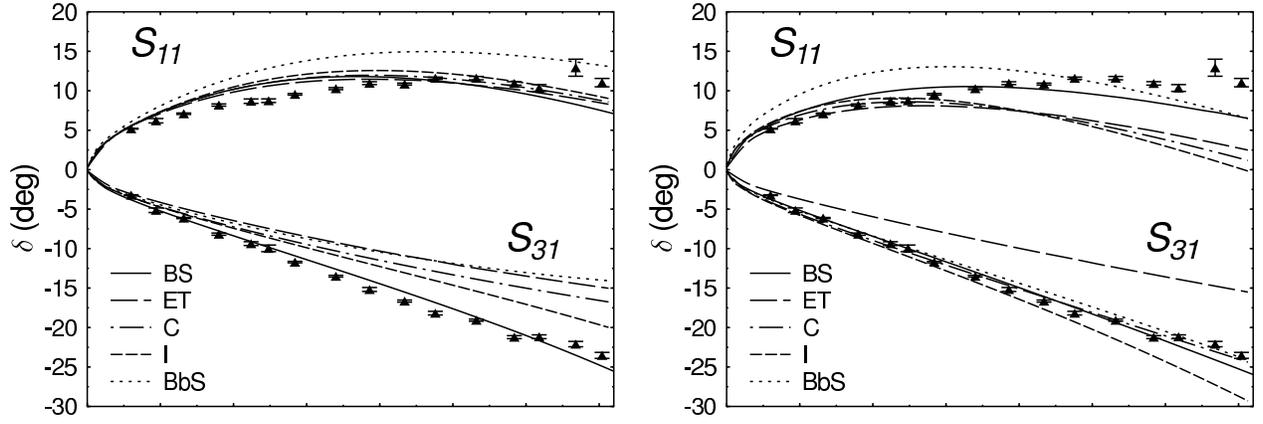}
\caption{ The $S_{11}$ and $S_{31}$ phase shifts for the  type II (a) and
type I (b) form factors. The data points are from
\cite{AS95}.}\label{Fig.6}
\end{figure}

In Fig.~\ref{Fig.6} we present the $S_{11}$ and $S_{31}$ phase shifts
for the two types of form factors and the different 3-D
approximations to the BS equation.
Also included are the results of the VPI SM95~\cite{AS95}
partial wave analysis. In this case it is
not clear as to which 3-D equation gives the best
approximation to the BS equation. This is partly due to the fact that
the final amplitude is not dominated by any single one of the diagrams that
contribute to the potential. In both the $S_{11}$ and $S_{31}$
partial waves there are large cancellations between the various
diagrams that contribute to the potential.

\begin{figure}[!]
   \centering\includegraphics[scale=1.0]{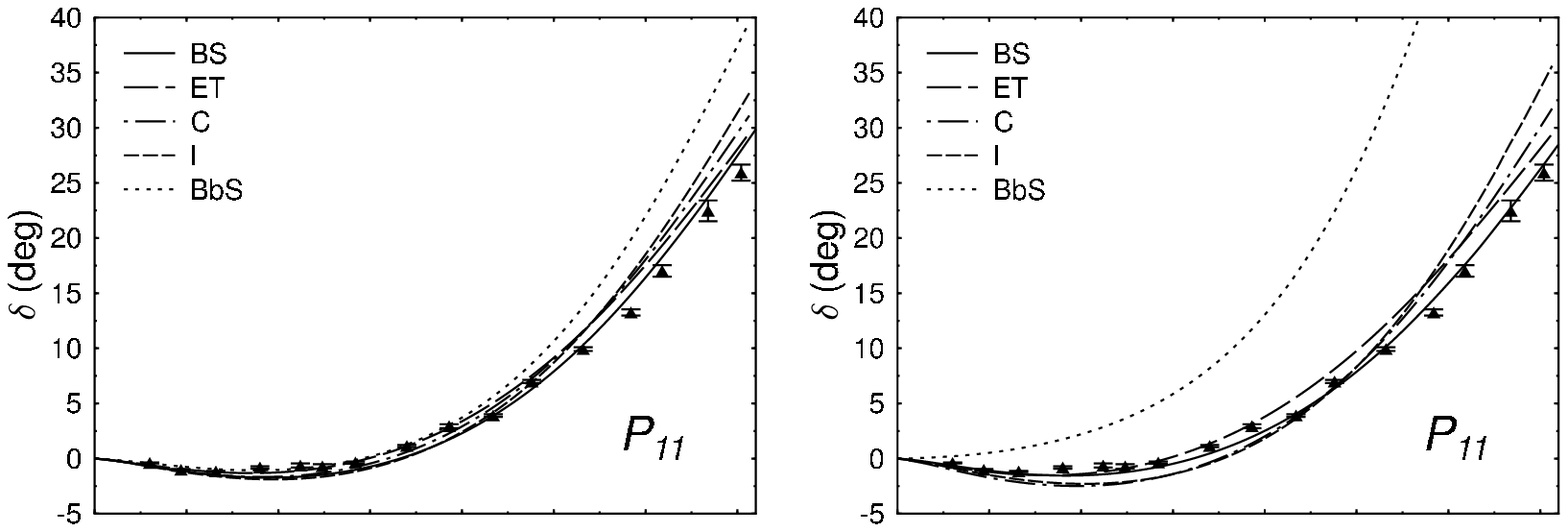}
\caption{ The $P_{11}$ phase shifts for the  type II (a) and
type I (b) form factors. The data points are from
\cite{AS95}.}\label{Fig.7}
\end{figure}

To see a  more pronounced difference between the different equations,
we turn next to the $P_{11}$ partial wave. In this case the amplitude
is dominated by the $s$-channel pole diagram, which is repulsive, as
well as the $u$-channel $\Delta$ pole, which is attractive.
In Fig.~\ref{Fig.7} we
present our results for the two different types of form factor. Here
it is clear that ET equation gives the best approximation to the BS
equation, but the variation between the different equations is not 
significant provided we restrict
our regularization to type II form factors that depend on the four-momentum 
of the pion only. For the type I form factor the 
BbS equation gives very poor results. At this stage we note
that the BbS equation gives additional attraction
when compared to the BS equation.

\begin{figure}[!]
   \centering\includegraphics[scale=1.0]{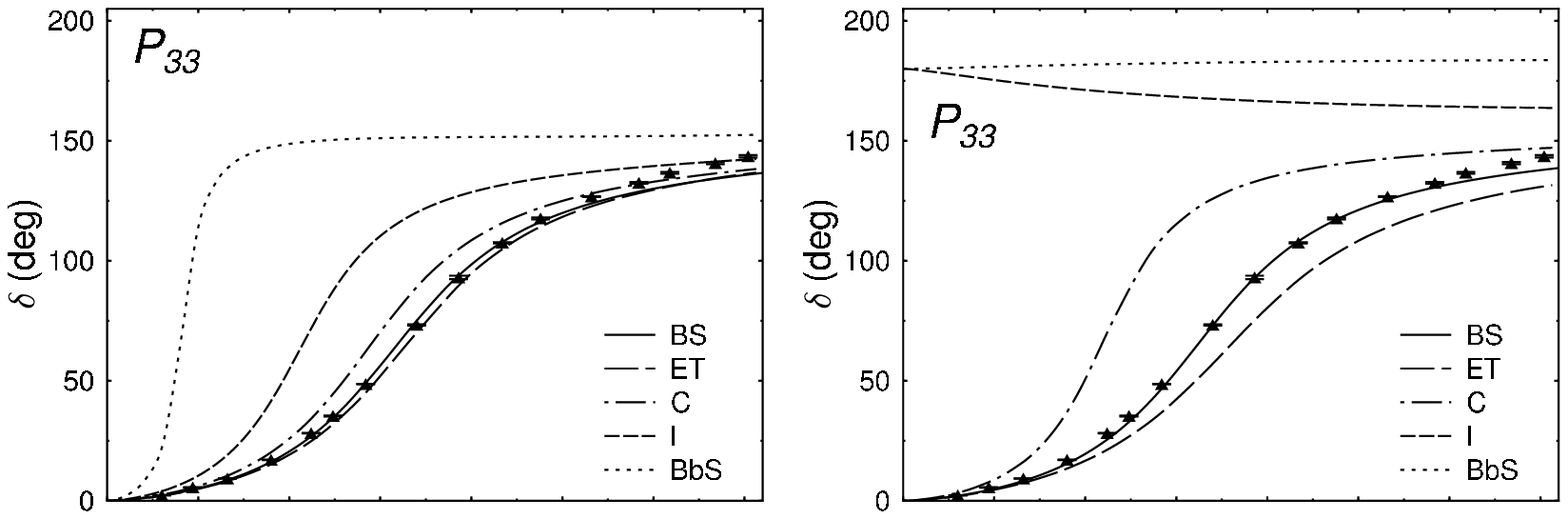}
\caption{ The $P_{33}$ phase shifts for the  type II (a) and
type I (b) form factors. The data points are from
\cite{AS95}.}\label{Fig.8}
\end{figure}

We finally turn to the $P_{33}$ partial wave which is dominated by
just one term in the potential: the $s$-channel $\Delta$
pole diagram. Here (see Fig.~\ref{Fig.8}) there is a dramatic change
as one goes from the BS equation to the ET equation to the Cohen
equation, and finally to the equations with just the two-body unitarity
thresholds. The trend in the phase shifts suggests that the
additional attraction generated by neglecting the higher unitarity cuts 
is slowly converting the resonance pole to a bound state pole. 
If it is the higher unitarity cuts that are the
source of this difference, then these cuts would be even more
important for higher-energy resonances in the $\pi N$ system. This is
a question that would require further investigation.

For the case of type I form factors, tables of the numerical
results are presented in Appendix~\ref{app.3}.

We do not present results for the inelasticities in the
$\pi N$ system, because the inelasticity generated by the BS equation 
and the different 3-D equations is negligible for the energy range
considered here.

\section{Conclusion}\label{sec.5}

In the preceding analysis we have demonstrated that the Bethe-Salpeter
equation in the ladder approximation has in addition to the two-body
unitarity cut, a selected set of $n$-body unitarity cuts. In reducing the
dimensionality of the BS equation from four to three it is possible to 
preserve some of the three- or four-body unitarity cuts and in this
way preserve some of the features of the original equation. This could 
be important at energies close to the three-body threshold, as is
the case for the $\pi N$ system in the region of the lowest resonances. From
the numerical results presented for the two models considered, we may
conclude that the preservation of these higher unitarity cuts do in
general result in an equation that is a better approximation to the BS 
equation. In particular, for the $\Delta$ resonance, the inclusion of
these higher thresholds are of considerable importance. 

It is always possible to readjust the parameters of the potential
within the framework of any one of these equations to fit the
experimental data. This procedure will result in the coupling constants
in the Lagrangian being different from those resulting from the use of 
the BS equation. As long as those coupling constants have little
physical meaning one may justify using any scattering equation that
preserves two-body unitarity. However, if we would like some 
consistency, e.g. between the $\pi N$ and $NN$ systems, then 
it may be important to reduce the approximations in the starting 
equation. More significant is the possibility of missing some 
physical mechanisms such as the contribution of three-body thresholds 
to resonances just above the pion production threshold.

Although three-body unitarity is not completely included in the ladder 
approximation to the BS equation, one could add further contributions
to three-body unitarity by dressing the nucleon in the $\pi N$
propagator. That will play an important role if the BS equation is to
include coupling to other channels such as the $\pi\Delta$ channel.

\acknowledgments

The authors would like to thank the Australian Research Council for
their financial support during the course of this work.
The work of A. D. L. was supported in part by a grant from the Natural
Sciences and Engineering Research Council of Canada.

\appendix

\section{Thresholds in the Equal-Time Equation}\label{app.1}

To illustrate the thresholds generated in the equal-time
formalism~\cite{K53,PW96,LA97}, we consider the singularities of
\begin{eqnarray}
\bra G\,V\,G\ket &=& \int\limits_{-\infty}^{+\infty}\,dq_{0}\
\frac{1}{\left(\frac{\sqrt{s}}{2}-q_{0}\right)^2 - E_q^2 +i\epsilon}\,
\frac{1}{\left(\frac{\sqrt{s}}{2}+q_{0}\right)^2 - E_q^2 
+i\epsilon}\\ \nonumber
&&\times\int\limits_{-\infty}^{+\infty}dq'_0\
\frac{1}{\left(q_0-q'_0\right)^2 - \omega^2_{q-q'} + i\epsilon}\,
\frac{1}{\left(\frac{\sqrt{s}}{2}-q'_{0}\right)^2 - E_{q'}^2 +i\epsilon}\,
\frac{1}{\left(\frac{\sqrt{s}}{2}+q'_{0}\right)^2 - E_{q'}^2 +i\epsilon}
                                                      \ .\label{eq:1a}
\end{eqnarray}
The singularity structure of $\bra G\,V\,G\ket$ is determined by the
pinching of the $q_0$ and $q'_0$ contours by the poles in the
integrand. For the $q'_0$ integration, we have poles from both the
propagators and the potential. As we have not partial wave expanded
the potential we have at this stage only poles. However, upon partial
wave expansion these poles will generate logarithmic branch points
which we will enumerate.

In the $q'_0$ plane we have the four propagator poles:
\begin{eqnarray}
1)\quad q'_0 &=& -\frac{1}{2}\sqrt{s} - E_{q'} + i\epsilon\ ;\quad\quad
2)\quad q'_0 = \frac{1}{2}\sqrt{s} - E_{q'} + i\epsilon\ ;\nonumber
\\
3)\quad q'_0 &=& -\frac{1}{2}\sqrt{s} + E_{q'} - i\epsilon\ ;\quad\quad
4)\quad q'_0 = \frac{1}{2}\sqrt{s} + E_{q'} - i\epsilon\ .
                                                      \label{eq:2a}
\end{eqnarray}
On the other hand the potential branch points in the $q'_0$-plane are at
\begin{eqnarray}
5)\quad q'_0 &=& q_0 - \omega_{q+q'} + i\epsilon\ ;\quad\quad
6)\quad q'_0 = q_0 - \omega_{q-q'} + i\epsilon\ ;\nonumber\\
7)\quad q'_0 &=& q_0 + \omega_{q+q'} - i\epsilon\ ;\quad\quad
8)\quad q'_0 = q_0 + \omega_{q-q'} - i\epsilon\ .\label{eq:3a}
\end{eqnarray}
For $\sqrt{s}>0$, the pinching of the $q'_0$ contour of integration is
between the following singularities:
\begin{itemize}
\item Between two of the poles of the propagator, and in particular
for $\sqrt{s}>0$, poles $2)$ and $3)$ pinch the contour provided
\begin{equation}
\sqrt{s} = 2E_{q'} - 2i\epsilon   \ .                  \label{eq:4a}
\end{equation}
In a similar manner for $\sqrt{s}<0$, poles $1)$ and $4)$ pinch the
contour to generate a branch cut at $\sqrt{s}=-2E_q+2i\epsilon$.
These will generate  thresholds at $\sqrt{s} =\pm 2m$.
\item Between one of the poles of the propagator and a branch point of
the potential. In particular singularities 3) and 6) pinch the
contour if
\begin{equation}
9)\quad q_0=-\frac{1}{2}\sqrt{s} + E_{q'} + \omega_{q-q'} - 2i\epsilon
\ .                                                  \label{eq:5a}
\end{equation}
This branch point is going to be involved in the pinching of the $q_0$
integration contour, and to that extent will not at this stage
generate a threshold in the final amplitude. Although the $q_{0}$
integration includes the on-shell point $q_{0}=0$, we do not expect
this singularity to generate a threshold in the amplitude, since we
can deform the $q_{0}$ contour of integration to avoid the point
$q_{0}=0$.

Similarly the propagator pole $2)$ and the branch point $8)$ pinch the
$q'_{0}$ contour provided
\begin{equation}
10)\quad q_0=\frac{1}{2}\sqrt{s} - E_{q'} - \omega_{q-q'} + 2i\epsilon
\ .                                                  \label{eq:6a}
\end{equation}
\end{itemize}
A similar set of branch points are generated for $\sqrt{s}<0$. In this
case the propagator poles $1)$ and $4)$ (see Fig.~\ref{Fig.1}) pinch
the $q_0'$ integration contour to produce a branch point at
$\sqrt{s}=-2m$. On the other hand the pinching of the
$q'_0$ by the propagator pole $1)$ and the potential branch point $8)$
generates a branch point in the $q_0$ plane at the point $q_0 =
-\frac{1}{2}\sqrt{s} - E_{q'} - \omega_{q-q'} + 2i\epsilon$. This is a
reflection of the branch point in Eq.~(\ref{eq:6a}) in the 
$\sqrt{s}$-plane. In a similar manner, we get a reflection in the
$\sqrt{s}$-plane of the branch point in Eq.~(\ref{eq:5a}) when
the pole $4)$ and the branch point $6)$ pinch the $q'_0$ contour. 
In this way the ET formulation maintains analyticity in $s$.

We now turn to the $q_0$ integration. In this case the singularities
of the integrand arise from the propagator poles and those generated by the
$q'_0$ integration. The propagator poles are at:
\begin{eqnarray}
11)\quad q_0 &=& -\frac{1}{2}\sqrt{s} - E_{q} + i\epsilon\ ;\quad
12)\quad q_0 = \frac{1}{2}\sqrt{s} - E_{q} + i\epsilon\ ;\nonumber
\\
13)\quad q_0 &=& -\frac{1}{2}\sqrt{s} + E_{q} - i\epsilon\ ;\quad
14)\quad q_0 = \frac{1}{2}\sqrt{s} + E_{q} - i\epsilon\ .
                                                      \label{eq:7a}
\end{eqnarray}
In this case the $q_0$ contour can be pinched by the following
singularities:
\begin{itemize}
\item For $\sqrt{s}>0$, propagator poles $12)$ and $13)$ pinch the
$q_{0}$ contour to give the branch cut
\begin{equation}
\sqrt{s} = 2E_q -2i\epsilon                      \ ,\label{eq:8a}
\end{equation}
while for $\sqrt{s}<0$ the poles $11)$ and $14)$ pinch the contour to
generate a branch cut at $\sqrt{s}= -2E_{q} +2i\epsilon$.
These will generate the two-body threshold $\sqrt{s}=\pm 2m$.
\item The propagator pole 12) and the branch cut generated by the
$q'_0$ integration 9) can pinch the $q_0$ contour. The condition for 
this pinch is
\begin{equation}
\sqrt{s} = E_q + E_q' + \omega_{q-q'} - 3i\epsilon\ .\label{eq:9a}
\end{equation}
Since this is in the kernel of the three-dimensional integral
equation, the actual pinch takes place at $q'=0$. As a result this
will generate a branch point at $\sqrt{s}=2m+\mu$.

In a similar manner the propagator pole 13) and the branch cut 
generated by the $q'_0$ integration 10) can pinch the $q_0$ contour. 
In this case the pinch takes place if
\begin{equation}
\sqrt{s} = E_q + E_q' + \omega_{q-q'} - 3i\epsilon\ , \label{eq:10a}
\end{equation}
which is identical to that in Eq.~(\ref{eq:9a}), and will generate the
three-body threshold at $\sqrt{s} = 2m+\mu$. A similar set of pinchings
of the $q_0$ contour takes place between the poles of the propagators
and the branch points generated by the $q'_0$ integration for
$\sqrt{s}<0$ to generate the three-body threshold at 
$\sqrt{s}= -(2m+\mu)$.
\end{itemize}
In this way we have established that the equal time equation generates
an amplitude that has two- and three-body thresholds and is analytic
in $s$. Unfortunately, not all possible three-body thresholds are
included, and therefore the resultant equation does not satisfy
three-body unitarity.

\section{3-D equations for $\pi N$ Scattering}\label{app.2}

For completeness, in this appendix we give explicit forms for the
different 3-D equations discussed in the present work. Since some 3-D equations
violate Lorentz invariance, for example the BbS equation, the 
solutions of such equations depend on
the choice of the relative four-momentum. Here we use the commonly-used choice
\begin{equation}
\alpha _N(s) = \frac{s + m_N^2 - \mu _{\pi}^2}{2s}  , \hspace*{1.0cm}
\alpha _{\pi}(s) = \frac{s + \mu _{\pi}^2 - m_N^2}{2s}  .
\end{equation}
In this section we make use of $\pi N \rightarrow \pi N$ amplitudes 
sandwiched between Dirac spinors.
We use the following notation for the amplitudes:
\begin{subequations}
\label{eq:sn12}
\begin{eqnarray}
T^{ \bar u u}(q',q;s) &=&\bar u
(\textbf{q}') T(q',q;s)u(\textbf{q})  ,\\
T^{ \bar v u}(q',q;s) &=&\bar v
(-\textbf{q}') T(q',q;s)u(\textbf{q})  ,
\end{eqnarray}
\end{subequations}
and similarly for the potentials. With this notation, the BS equation has
the form of two coupled equations for $T^{ \bar u u}$ and $T^{ \bar v u}$~\cite{LA99}:
\begin{equation}
T^{ \bar w u}(q',q;P) =   V^{ \bar w u}(q',q;P) 
 - {i \over (2 \pi)^4} \sum _{w''=u,v} 
\int d^4 q'' \, V^{ \bar w w''}(q',q'';P)
G^{\bar w'' w''}_{\pi N}(q'';P)T^{ \bar w'' u}(q'',q;P)  ,
\end{equation}
with $w=u \mbox{, } v$. The $\pi N$ propagators $G^{\bar w'' w''}_{\pi N}$ 
are defined in Eq.~(\ref{eq:bsegf}).

Firstly we consider the Cohen equation, which has the form (before partial 
wave expansion)
\begin{equation}
T^{ \bar w u}(\textbf{q}',\textbf{q};s) = 
V^{ \bar w u}(0,\textbf{q}';0,\textbf{q};s)
 - {i \over (2 \pi)^4} \sum _{w''=u,v}
\int d \textbf{q}'' \, \tilde K^{ \bar w w''}(\textbf{q}',\textbf{q}'';s)
T^{ \bar w'' u}(\textbf{q}'',\textbf{q};s)  , \label{eq:coh}
\end{equation}
with the kernel given by
\begin{equation}
 \tilde K^{ \bar w' w}(\textbf{q}',\textbf{q};s) = 
 \int _{- \infty} ^{\infty} dq_0  V^{ \bar w' w}(0,\textbf{q}';q_0,\textbf{q};s)
 G_{\pi N} ^{\bar w w} (q_0,q;s) .
 \end{equation}
 The relative energy integration can be carried out by making use
 of the Wick rotation~\cite{Wi54}.

The remaining 3-D equations for $\pi N$ scattering have the form 
\begin{equation}
T^{ \bar w u}(\textbf{q}',\textbf{q};s) = 
K^{ \bar w u}(\textbf{q}',\textbf{q};s) 
 - {i \over (2 \pi)^4} \sum _{w''=u,v}
\int d \textbf{q}'' \, K^{ \bar w w''}(\textbf{q}',\textbf{q}'';s)
g^{\bar w'' w''}_{\pi N}(q'';s)T^{ \bar w'' u}(\textbf{q}'',\textbf{q};s) . \label{eq:3deqn}
\end{equation}
The two-body $\pi N$ propagators are denoted as 
$g^{\bar u u}_{\pi N}$ and $g^{\bar v v}_{\pi N}$, which contain the positive and negative energy
components of the nucleon propagator, respectively. The 3-D potentials are denoted
as $K ^{\bar w' w}$.

To our knowledge the ET equation for $\pi N$ scattering has not been 
previously discussed in the literature.
The two-body $\pi N$ propagator appearing in
the ET equation is obtained from the propagator in the BS equation
by integrating out the relative energy, i.e.,
\begin{eqnarray}
g_{\pi N}(\textbf{q};s) & = &
\int_{- \infty}^{\infty} dq_0 G_{\pi N} (q_0, \textbf{q};s) \nonumber \\
& = & 2 \pi i {m_N \over 2 E_q \omega _q}
\left( { \Lambda ^+(\textbf{q}) \over
\sqrt{s} - E_q - \omega _q + i \epsilon} +
{ \Lambda ^- (- \textbf{q}) \over
\sqrt{s} + E_q + \omega _q - i \epsilon} \right) \nonumber \\
& \equiv & g_{\pi N}^{\bar u u}(q;s) \Lambda ^+(\textbf{q}) +
g_{\pi N}^{\bar v v}(q;s) \Lambda ^-(- \textbf{q}) . \label{eq:1b}
\end{eqnarray}
For the calculation of the ET potential, we will also need the inverse
of $g_{\pi N}$, which is given by
\begin{eqnarray}
g_{\pi N}(\textbf{q};s)^{-1} =
{1 \over 2 \pi i} {2 m_N \omega _q \over E_q }
\Biggl( [\sqrt{s} - E_q - \omega _q] \Lambda ^+ (- \textbf{q}) +
[\sqrt{s} + E_q + \omega _q] \Lambda ^- (\textbf{q})  \Biggr)\ . \label{eq:2b}
\end{eqnarray}
The potential for the ET equation is then
\begin{eqnarray}
K (\textbf{q}',\textbf{q};s) & = &
g_{\pi N}(\textbf{q}';s)^{-1}
\int_{- \infty} ^{\infty}dq'_0 \int_{- \infty} ^{\infty}dq_0
G_{\pi N}(q'_0,\textbf{q}';s) V(q'_0,\textbf{q}';q_0,\textbf{q};s)
G_{\pi N}(q_0,\textbf{q};s) \nonumber \\
& & \times g_{\pi N}(\textbf{q};s)^{-1} .          \label{eq:3b}
\end{eqnarray}
Multiplying Eq.~(\ref{eq:3b}) from the left and right by Dirac spinors, 
and making use of the orthonormality of Dirac spinors, we find 
\begin{eqnarray}
\hspace*{-0.9cm} K^{\bar w' w} (\textbf{q}',\textbf{q};s)& = & - {1 \over (2 \pi)^2}
  {2 E_{q'} \omega _{q'} \over m_N}
  {2 E_{q} \omega _{q} \over m_N}
\left(\sqrt{s} - \xi ' E_{q'} - \xi '\omega _{q'} \right)
\left(\sqrt{s} - \xi E_{q} - \xi \omega _{q} \right) \nonumber \\
& & \times \int _{- \infty}^{\infty} dq'_0 \int _{- \infty}^{\infty} dq_0
G_{\pi N}^{\bar w' w'}(q'_0,q';s) V^{\bar w' w}(q'_0,
\textbf{q}';q_0,\textbf{q};s)
  G_{\pi N}^{\bar w w}(q_0,q;s) .  \label{eq:7b}
\end{eqnarray}
Here $\xi ' = +1$ ($-1$) for $w'=u$ ($v$), and
$\xi  = +1$ ($-1$) for $w=u$ ($v$).
Again, the relative energy integrations above can be performed numerically by
making use of the Wick rotation~\cite{Wi54}, which can be applied in
the same way as it is used in solving the BS equation~\cite{LA99}.

For the instantaneous equation, the $g^{\bar w w}_{\pi N}$ are obtained by integrating
out the relative energy of the $\pi N$ propagator appearing in the
BS equation, and so we have
\begin{subequations}
\begin{eqnarray}
g^{\bar u u}_{\pi N}(q;s) & = & { \pi i m_N \over E_q \omega _q 
                          [ \sqrt{s} - E_q - \omega _q + i \epsilon ]}, \\			  
g^{\bar v v}_{\pi N}(q;s) & = &{ \pi i m_N \over E_q \omega _q 
                          [ \sqrt{s} + E_q + \omega _q - i \epsilon  ]}.
\end{eqnarray}	
\end{subequations}
It is clear that the ET and I equations make use of the same $\pi N$
propagator.

The BbS equation is obtained by replacing $G_{\pi N}$ in the BS equation
with the following two-body propagator
\begin{equation}
G^{\rm BbS}_{\pi N}(q;P) = 
2 \pi i \int_{s_{th}}^{\infty}
\frac{ds' }{s' - s - i \epsilon}  
 [ \alpha _N \slasss{P} \, ' + \slass{q} + m_N] 
  \delta ^{(+)}[(\alpha _N P'+q)^2-m_N^2] 
  \delta ^{(+)}[(\alpha _{\pi} P'-q)^2-\mu _{\pi}^2]  ,
\end{equation}
where $s_{th} = (m_N + \mu _{\pi})^2$, and $P'=(\sqrt{s'}/ \sqrt{s}) P$. This 
results in the BS equation being reduced to a 3-D equation of the form
of Eq.~(\ref{eq:3deqn}), with the $\pi N$ propagators given by
\begin{subequations}
\begin{eqnarray}
g^{\bar u u}_{\pi N}(q;s) & = & { 2 \pi i m_N \over E_q \omega _q 
                          [ s -( E_q + \omega _q )^2 + i \epsilon ]}, \\			  
g^{\bar v v}_{\pi N}(q;s) & = & 0.
\end{eqnarray}	
\end{subequations}
Notice that the coupling to negative energy states is neglected, but 
$g^{\bar u u}_{\pi N}$ has poles at both $\sqrt{s} = E_q + \omega_q$ and
$\sqrt{s} = -E_q - \omega _q$.

In the I equation
the potential is assumed to be static, while for the
BbS equation the relative energy is fixed at its on-shell value.
Therefore the potentials for these two equations are given by
\begin{equation}
K^{ \bar w' w}(\textbf{q}',\textbf{q};s) = 
V^{ \bar w' w}(0,\textbf{q}';0,\textbf{q};s) .
\end{equation}
The differences between the I and BbS equations lie in the 
form used for the $\pi N$ propagator.

\section{Tabulated Results}\label{app.3}

Tables~\ref{tab.res1} to \ref{tab.res4} show the numerical values of the
$\pi N$ phase shifts when type I form factors are used.

\bibliography{bs3d}


\begin{table}
\begin{center}
\begin{tabular}{ddddddddd}\hline
E_{\mbox{\scriptsize lab}} & \mbox{BS} & \mbox{ET} & \mbox{C} & \mbox{I} & 
\mbox{BbS}  \\ \hline
 50 &  6.78 &  6.30 &  6.34 &  7.34  &  9.33  \\
100 &  9.17 &  7.84 &  7.96 &  9.18  & 12.9   \\
150 & 10.2  &  8.05 &  8.12 &  8.76  & 13.0   \\
200 & 10.3  &  7.42 &  7.35 &  7.39  & 12.3   \\
250 &  9.76 &  6.21 &  5.92 &  5.40  & 11.0   \\
300 &  8.72 &  4.62 &  4.06 &  2.95  &  9.13  \\
350 &  7.18 &  2.74 &  1.90 &  0.191 &  6.85  \\ \hline
\end{tabular}
\end{center}
\caption{The $S_{11}$ phase shifts using the type I form factors as calculated
by the BS equation and the different 3-D quasipotential equations.}
\label{tab.res1}
\end{table}

\begin{table}
\begin{center}
\begin{tabular}{ddddddddd}\hline
E_{\mbox{\scriptsize lab}} & \mbox{BS} & \mbox{ET} & \mbox{C} & \mbox{I} & 
\mbox{BbS}  \\ \hline
 50 &  -5.46 &  -3.73 &  -5.93 &  -5.52 &  -5.21   \\
100 &  -8.73 &  -5.96 &  -9.24 &  -9.34 &  -8.45   \\
150 & -11.9  &  -8.02 & -12.3  & -13.0  & -11.4    \\
200 & -15.0  &  -9.96 & -15.1  & -16.5  & -14.4    \\
250 & -18.2  & -11.8  & -18.0  & -20.2  & -17.4   \\
300 & -21.4  & -13.6  & -20.7  & -24.3  & -20.5    \\
350 & -24.9  & -15.3  & -23.5  & -28.7  & -23.7    \\ \hline
\end{tabular}
\end{center}
\caption{The $S_{31}$ phase shifts using the type I form factors as calculated
by the BS equation and the different 3-D quasipotential equations.}
\label{tab.res2}
\end{table}

\begin{table}
\begin{center}
\begin{tabular}{ddddddddd}\hline
E_{\mbox{\scriptsize lab}} & \mbox{BS} & \mbox{ET} & \mbox{C} & \mbox{I} & 
\mbox{BbS}  \\ \hline
 50 & -1.09 &  -1.21  &  -1.50 &  -1.07 &  0.266  \\
100 & -1.53 &  -1.40  &  -2.46 &  -2.34 &  1.65   \\
150 & -0.50 &   0.159 &  -1.36 &  -1.52 &  4.60   \\
200 &  2.45 &   3.793 &   2.03 &   1.62 & 10.9    \\
250 &  7.75 &   9.78  &   8.26 &   7.95 & 23.6    \\
300 & 15.7  &  18.1   &  17.7  &  18.4  & 45.2    \\
350 & 25.9  &  28.10  &  20.7  &  32.7  & 70.9    \\ \hline
\end{tabular}
\end{center}
\caption{The $P_{11}$ phase shifts using the type I form factors as calculated
by the BS equation and the different 3-D quasipotential equations.}
\label{tab.res3}
\end{table}

\begin{table}
\begin{center}
\begin{tabular}{ddddddddd}\hline
E_{\mbox{\scriptsize lab}} & \mbox{BS} & \mbox{ET} & \mbox{C} & \mbox{I} & 
\mbox{BbS}  \\ \hline
 50 &   4.85 &   4.61 &   8.23 & 174.0 & 180.0   \\
100 &  18.7  &  16.7  &  45.4  & 171.0 & 181.0   \\
150 &  51.6  &  42.2  & 111.0  & 169.0 & 182.0   \\
200 &  96.0  &  80.5  & 133.0  & 167.0 & 183.0  \\
250 & 120.0  & 108.0  & 141.0  & 165.0 & 183.0   \\
300 & 131.0  & 121.0  & 144.0  & 164.0 & 183.0   \\
350 & 138.0  & 131.0  & 147.0  & 163.0 & 183.0   \\ \hline
\end{tabular}
\end{center}
\caption{The $P_{33}$ phase shifts using the type I form factors as calculated
by the BS equation and the different 3-D quasipotential equations.}
\label{tab.res4}
\end{table}

 
 \newpage
 
\newpage

\newpage

\newpage

\newpage

\newpage

\newpage

\newpage

\end{document}